# Full frame denoising for pyramid wavefront sensors

Noah Schwartz*[a,c], Samuel Pinilla[b], Charlotte Z. Bond[a], Ben Buky[a], Gianfelice Cinque[b], Robert Rambo[b]

[a]UK Astronomy Technology Centre, Blackford Hill, Edinburgh EH9 3HJ, United Kingdom; [b]Diamond Light Source, Chilton, Didcot, United Kingdom; [c]Marseille Univ, CNRS, LAM, Laboratoire d'Astrophysique de Marseille, Marseille, France.

## ABSTRACT

Adaptive optics systems operating under low-flux conditions face significant challenges, as photon and detector noise in particular degrade wavefront measurements and ultimately limit correction performance. While pyramid wavefront sensors (PyWFSs) offer greater sensitivity than conventional wavefront sensors such as the Shack-Hartmann sensor in many operating regimes, obtaining accurate wavefront estimates under photon-starved conditions remains a key challenge. We present a full-frame image denoising strategy for the PyWFS that exploits the nonlocal self-similarity of wavefront sensor image patches through FFT-accelerated patch grouping and global collaborative 3D wavelet filtering, applied directly to the raw PyWFS intensity frame prior to slope computation. Specifically, the method suppresses noise while preserving structural features required for accurate wavefront reconstruction.

The approach is evaluated using end-to-end simulations of a VLT-scale SCAO system. The results show improved performance in low signal-to-noise regimes, with typical Strehl ratio gains of up to 12% and an increase in limiting magnitude of approximately 0.5 in median seeing conditions. Modal analysis indicates reduced variance across most controlled modes. The improved PSF quality enables a reduction in the FWHM and an enhanced contrast. These results demonstrate that image-domain denoising can improve the robustness of PyWFS-based AO systems and extend their operational range toward fainter guide stars.

## 1 INTRODUCTION

Pyramid wavefront sensors (PyWFS) are widely used in high-performance adaptive optics (AO) systems, including instruments such as HARMONI [1, 2] for the ELT and future instruments such as the ELT's Planetary Camera and Spectrograph (PCS) [3] for high-contrast imaging. While the PyWFS offers high sensitivity in closed loop, its performance degrades significantly under low-flux conditions due to photon and detector noise, limiting wavefront measurement accuracy and achievable Strehl ratio. At faint guide star magnitudes, noise dominates the PyWFS signal, impacting wavefront estimation, loop stability, and modal control. Existing mitigation strategies, such as temporal filtering, provide only partial improvements and do not directly address noise at the WFS frame level. As a result, extending AO performance toward lower signal-to-noise regimes remains a key challenge.

In this work, we investigate a full-frame denoising strategy applied directly to PyWFS images prior to slope computation, and evaluate its potential to reduce modal noise propagation, extend the controllable modal basis, and improve overall AO performance without introducing reconstruction bias or compromising loop stability. The method is based on 3D sparse representations in the transform domain, grouping similar patches to suppress noise while preserving structural information. This enables improved noise propagation through the AO loop without modifying the reconstruction framework.

*noah.schwartz@stfc.ac.uk

The performance of the proposed approach is evaluated through end-to-end simulations with the OOPAO framework for a VLT-scale SCAO system. We assess the impact on residual wavefront error and Strehl ratio, with particular emphasis on low-flux regimes. The results demonstrate enhanced robustness and highlight the potential of advanced denoising techniques to extend the operational limits of pyramid-based AO systems.

# 2 DENOISING ALGORITHM

## 2.1 Description of the algorithm

The proposed denoising approach is rooted in the principle of nonlocal self-similarity: natural images - and, by extension, wavefront sensor intensity frames - contain recurring structures across spatially separated regions. By identifying and grouping similar image patches into three-dimensional (3D) stacks, it becomes possible to perform collaborative filtering that exploits these redundancies, suppressing noise far more effectively than methods that operate on individual pixels or local neighbourhoods alone [6].

The algorithm operates on the full PyWFS detector frame (see Figure 1) encompassing all four pupil images simultaneously. This choice is deliberate: treating the four pupils as a single image rather than processing each pupil in isolation makes the patch-grouping step aware of the structural correlations across pupils at any given wavefront state, thereby increasing the pool of mutually similar patches available for collaborative filtering. It is important to note that at this stage, the algorithm does not make use of the information between the pupils (which is mostly noise) to infer any information on the noise characteristics.

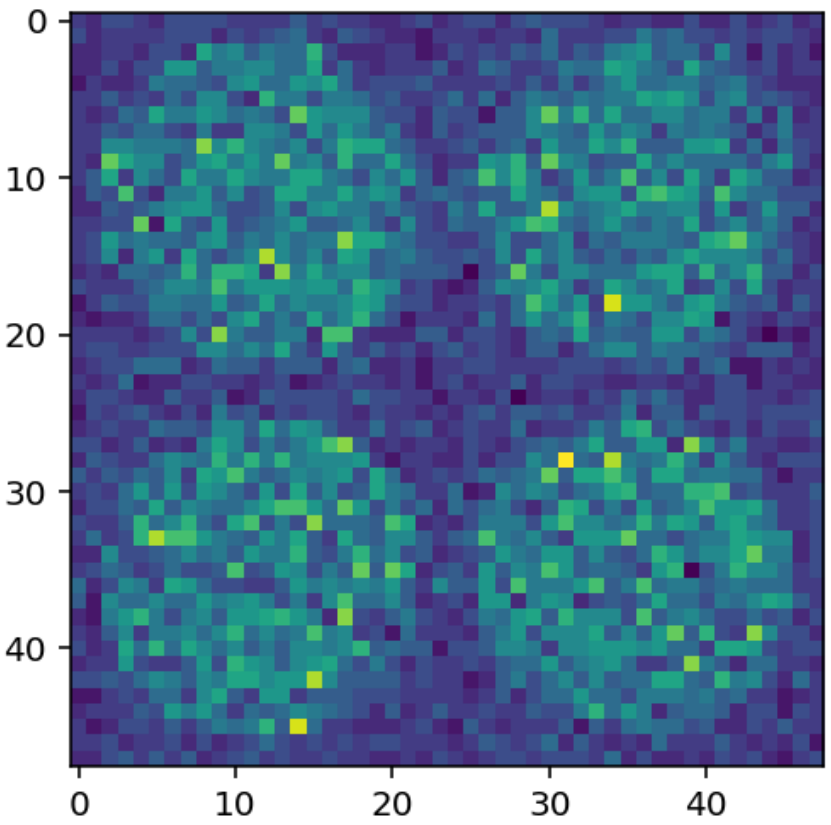

**Figure 1: Illustration of typical full Pyramid WFS detector frame. The footprint of the telescope pupil and the noise is clearly visible. The frame is 48x48 pixels.**

**Block matching.** The first stage identifies, for each reference patch in the image, the set of most similar patches within a local search window. Given a reference block $z$ of size $N \times N$, the squared Euclidean distance to any candidate block $z^{kl}$ is decomposed as:

$$\left\|z^{ij} - z^{kl}\right\|^2 = \left\|z^{ij}\right\|^2 + \|z^{kl}\|^2 - 2\langle z^{ij}, z^{kl}\rangle$$

The two squared-norm terms can be precomputed for all possible blocks in a single pass over the image, yielding a stored norm map. For each reference block, the remaining inner-product term is then evaluated as a cross-correlation via the Fast Fourier Transform (FFT) within a local search window of size $M_{loc}$. This reduces the dominant cost of block matching - which in classical implementations requires a large number of independent patch comparisons - to a single FFT-based cross-correlation per reference block, giving a substantial computational saving [6]. Reference blocks tile the image without overlap, and for each reference block the $K = 16$ nearest neighbours within the search window are retained.

**Global 3D wavelet thresholding.** Rather than filtering each matched group of patches independently, the matched blocks from all reference blocks are assembled jointly into a single 3D volume $V$ of spatial dimensions equal to the full image and depth $K$. The first slice of this volume contains the noisy image; the matched blocks for each reference block are stacked in the depth dimension at the spatial coordinates of that reference block. A single filtered 3D orthonormal wavelet transform is then applied to the entire volume $V$ ::

$$U = T_{3D}^{-1}\left(Y_{\alpha\sigma}\left(T_{3D}(V)\right)\right)$$

where $Y_{\alpha\sigma}$ denotes hard thresholding at a level $\alpha$ proportional to the estimated noise standard deviation $\sigma$. This global formulation is substantially more computationally efficient than the per-group approach, because large-scale FFT operations on a single volume benefit far more from hardware acceleration than many small independent transforms [6]. The denoised blocks are aggregated back into a 2D image using weights inversely proportional to the 3D total variation of each matched group, following the procedure of [6].

To reduce boundary artefacts and improve the statistical quality of the estimate, two complementary averaging strategies are employed. First, translation-invariant cycle-spinning is applied: the volume is filtered twice, with a circular shift applied before the second pass, and the two estimates are averaged - a well-established technique in wavelet denoising [7]. Second, the entire procedure (block matching, volume assembly, filtering, and aggregation) is repeated after translating the input image by $\frac{N}{4}$ and $\frac{N}{2}$ pixels, with the results of all translations averaged to produce the final denoised frame. These repeated passes are run in parallel and do not materially increase wall-clock time [6].

**Noise estimation and adopted configuration.** The noise standard deviation $\sigma$ required by the thresholding step is estimated directly from each incoming PyWFS frame using an adaptive multilayer wavelet decomposition approach. Following Donoho's theory [7], when an image is subjected to wavelet decomposition the noise coefficients are distributed evenly across scales, while the amplitude of signal coefficients decreases with increasing scale. Crucially, the characteristic distributions of noise and signal in the low-frequency wavelet sub-bands become progressively more similar at deeper decomposition levels. The estimation procedure therefore decomposes the incoming frame through multiple wavelet layers, applies a Principal Component Analysis (PCA)-based variance estimator to the approximate representation in the first few low-frequency sub-bands, and synthesises the per-layer estimates into a single noise variance value. To avoid instability caused by the PCA estimator alone, the method also incorporates local block variance statistics to cross-validate the appropriate decomposition depth: a layer is discarded when the ratio of adjacent-layer coefficient amplitudes exceeds the robust attenuation threshold of 0.6745 [7], which is consistent with the Donoho robust estimator $\frac{\sigma=\text{median}(W_{hh})}{0.6745}$.

When the estimated noise variance falls below the threshold corresponding to this attenuation value, the PCA-based estimate alone is used directly. In practice, the resulting estimate is scaled by a tunable factor α to account for the specific signal-to-noise characteristics of the PyWFS image, so that the effective denoising threshold is proportional to $\sigma$. The optimal value of $\sigma$ depends on guide star magnitude and observing conditions and is discussed in detail in Section 4.2. At each iteration of the AO control loop, the full PyWFS frame is passed through the noise estimator and the denoising algorithm in sequence, and the denoised frame is used directly for slope computation without any modification to the downstream reconstruction pipeline.

### 2.2 Performance

**Hardware scaling, and execution time.** On CPU, the dominant computational cost is the FFT-based cross-correlation block matching, which scales as $O(M^2 \log M)$ with the number of pixels $M^2$. For GPU execution, the global 3D wavelet thresholding step - consisting entirely of convolution-type operations - maps naturally onto batched FFT kernels and benefits strongly from the parallelism of GPU architectures. The execution time of denoising a PyWFS frame is approximately 0.79s on CPU hardware. GPU deployment constitutes an alternative path to deterministic low-latency execution: the block matching and hard-thresholding stages are dataflow-friendly and amenable to pipelined fixed-point arithmetic, making GPU an attractive target for systems where latency guarantees are critical. Characterisation of GPU performance for this application is identified as future work.

## 3 VLT-SCALE SIMULATION

### 3.1 SCAO simulation conditions

In this paper, we evaluate the performance of a VLT-scale single-conjugate adaptive optics (SCAO) system through end-to-end simulations performed with OOPAO (Object-Oriented Python Adaptive Optics toolbox) [4]. The principal simulation parameters are summarised in Table 1. The nominal loop frequency is 1 kHz, with additional simulations conducted at 500 Hz and 250 Hz to assess temporal performance. The control strategy is based on a simple integrator with a two-frame delay and a fixed loop gain of 0.4. Unless otherwise specified, the pyramid wavefront sensor (PyWFS) detector read noise is set to 1 $e^-$. The analysis presented here is restricted to the atmospheric correction performance of the AO loop. Additional error terms - such as calibration uncertainties, non-common path aberrations, and structural or vibrational disturbances - are not considered.

**Table 1: Summary of the nominal AO simulation parameters used.**

| Atmospheric parameters | | Pyramid WFS | | Deformable Mirror | |
|---|---|---|---|---|---|
| $r_0$ @ 500 nm | 0.15 m | Pupil diameter | 20 px | Number of actuators | 356 |
| $L_0$ | 25.0 m | Pupil separation | 4 px | Pitch | 0.38 m |
| $Tau_0$ | 0.0037 s | Modulation radius | 3 λ/D | Mechanical coupling | 35% |
| $V_0$ | 12.55 m/s | Post-processing | slopes | Misregistration | None |

The PyWFS signals are processed in terms of slope measurements, analogous to the x- and y-direction gradients obtained with classical wavefront sensors such as the Shack-Hartmann WFS. A quad-cell formulation is applied to the four pupil images to generate two slope maps. Here, $I_n$ denotes the intensity map of each of the four PyWFS pupil images ($n = 1,2,3,4$), and $I_{\text{norm}}$ represents the total integrated intensity across all four pupil images:

$$S_X = \frac{I_1 + I_3 - I_2 - I_4}{I_{norm}}, S_Y = \frac{I_1 + I_2 - I_3 - I_4}{I_{norm}}$$

The simulated atmosphere is composed of 5 layers respectively at an altitude of {0, 1, 5, 10, 20}km, a fractional Cn2 of {45, 10, 10, 25, 10}% and wind speeds of {10, 12, 11, 15, 20}m/s. The deformable mirror (DM) has 21x21 actuators with 0.35 actuator coupling.

To optimise performance, the SCAO control loop operates on a truncated set of Karhunen–Loève (KL) modes, with the number of controlled modes adapted to the observing conditions (i.e. loop frequency, PyWFS read-out-noise, $r_0$…). For the nominal configuration, the adopted modal truncation is summarised in Table 2.

In this analysis, the same modal basis and truncation are retained for both the noisy and denoised cases. Consequently, the denoised case is not strictly optimised, and its performance should be regarded as conservative. In particular, the improved noise characteristics would likely allow stable control of additional higher-order modes, suggesting that further optimisation could yield additional performance improvements.

**Table 2: Number of KL modes controlled as a function of stellar magnitude and sampling frequency.**

| **Mag.** | **10** | **12** | **13** | **14** | **14.5** | **15** | **15.5** | **16** | **16.5** | **17** |
|---|---|---|---|---|---|---|---|---|---|---|
| **250 Hz** | 300 | 275 | 270 | 220 | 200 | 175 | 130 | 90 | 60 | 40 |
| **500 Hz** | 340 | 340 | 271 | 250 | 200 | 130 | 90 | 60 | 40 | 20 |
| **1 kHz** | 340 | 330 | 225 | 125 | 100 | 70 | 40 | 20 | 10 | 10 |

### 3.2 Nominal simulation results

Figure 2 presents the closed-loop pyramid wavefront sensor (PyWFS) image (one quadrant displayed) as a function of stellar magnitude, assuming a detector readout noise of 1 $e^-$ and a frame rate of 1 kHz. At this flux level, the PyWFS receives approximately 353 photons per frame for a magnitude 15. As a consequence, the signal is dominated by noise, and the pupil structure becomes virtually indistinguishable.

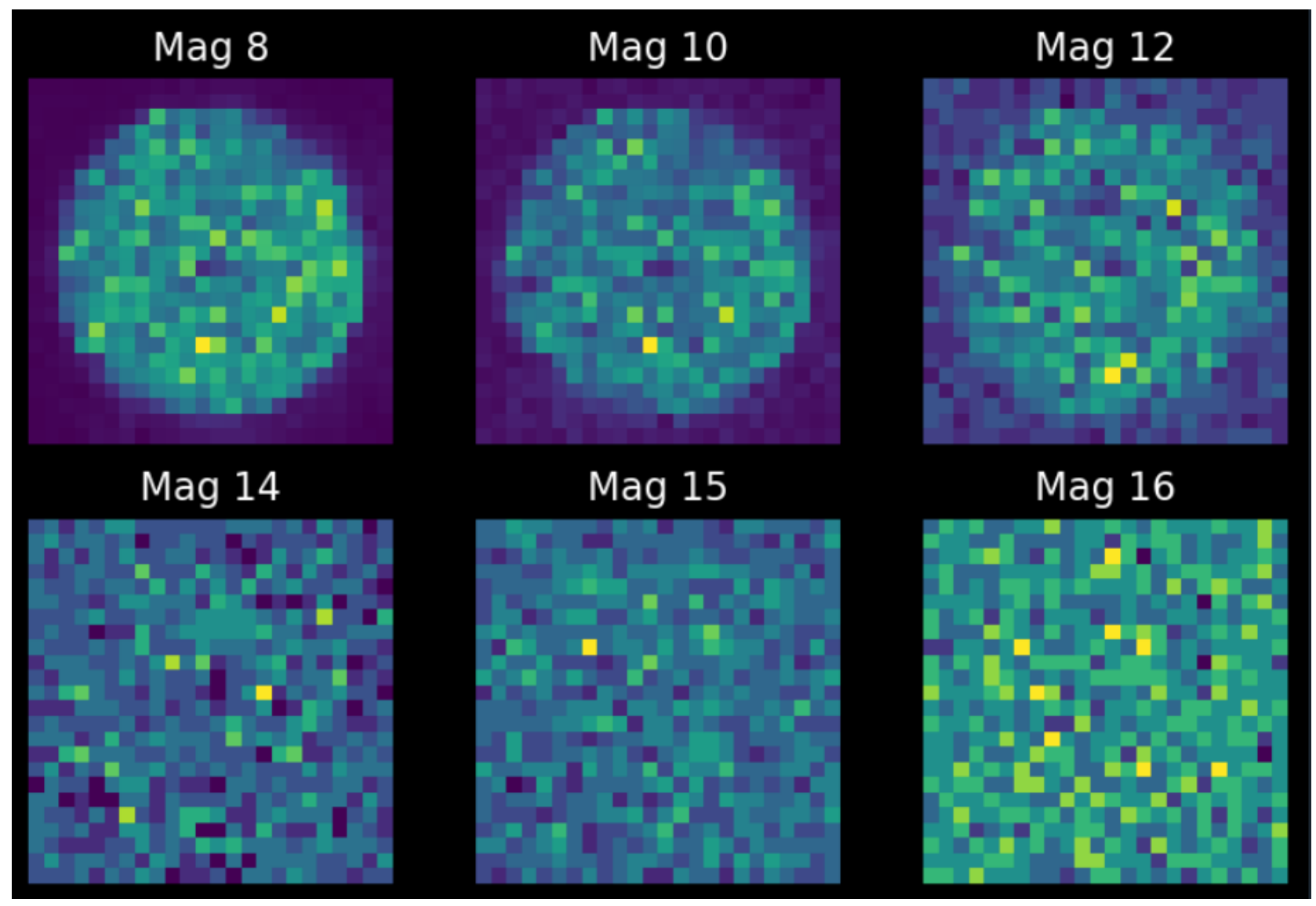

**Figure 2: Closed loop pyramid frame image (only one quadrant is shown) with noise for different magnitudes. Wavefront sensing in I-band. RON = 1e$^-$. Frame rate =1 kHz.**

The results presented in this paper are expressed as a function of the natural guide star magnitude. Table 3 reports the corresponding photon flux incident on the full PyWFS detector for each magnitude.

**Table 3: Number of photons on the PyWFS (full frame) as a function of stellar magnitude.**

| Magnitude | 10 | 12 | 13 | 14 | 15 | 16 | 17 | 18 |
|---|---|---|---|---|---|---|---|---|
| #Photons [1 kHz] | 35308 | 5596 | 2228 | 887 | 353 | 141 | 56 | 22 |

Figure 3 (left) presents the residual wavefront error as a function of iteration for different guide star magnitudes at a sampling frequency of 1 kHz. The impact of photon noise, driven by guide star magnitude, becomes increasingly evident as the flux decreases. The figure also shows (right) the Strehl ratio (SR) as a function of guide star magnitude for different frame rates. A limiting magnitude of approximately 18 is observed for a frame rate as low as 250 Hz. Naturally, reducing the frame rate further (e.g., to 125 Hz) may improve sensitivity by increasing the photon flux per frame, potentially extending the limiting magnitude slightly. For a magnitude 10 star, the closed-loop performance gives a residual wavefront error of 103.4 nm RMS, or a SR = 91.6% (resp. 86.9% and 62.2% for sampling frequency of 500 Hz and 250 Hz).

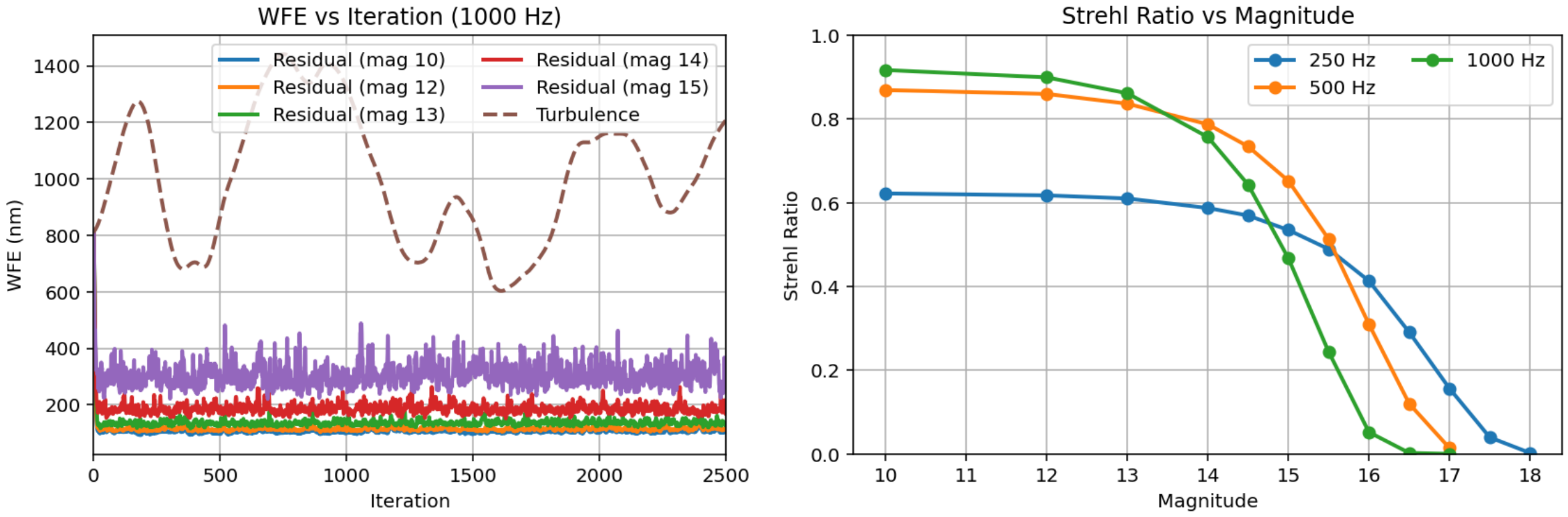

**Figure 3: (Left) Residual wavefront error as a function of iteration and guide star magnitude for a sampling frequency of 1 kHz. (Right) Strehl ratio as a function as guide star magnitude and frame rate.**

# 4 DENOISED SIMULATIONS

Denoising is one of several complementary approaches for improving AO performance in low-SNR conditions. In this work, we optimise the number of controlled KL modes to avoid correcting noise-dominated high-order modes, and we investigate the impact of the PyWFS modulation radius, which governs the trade-off between sensitivity and robustness.

## 4.1 General performance: Strehl Ratio

Figure 4 presents the improvement provided by the denoising algorithm. At each iteration, the full PyWFS frame is passed through the denoising algorithm. The slopes are then computed using the denoised frame directly. It is important to note that the same number of KL modes are controlled for the noisy and denoised cases (see Table 2), leading to potential further optimisation and a potential improved performance for the denoised case.

For bright star case (limited by photon noise) the denoising does not improve over the nominal case. When the performance starts to degrade, the denoising clearly outperforms the noisy case with gains reaching up to +12% SR. The limiting magnitude is increased by up to +0.5.

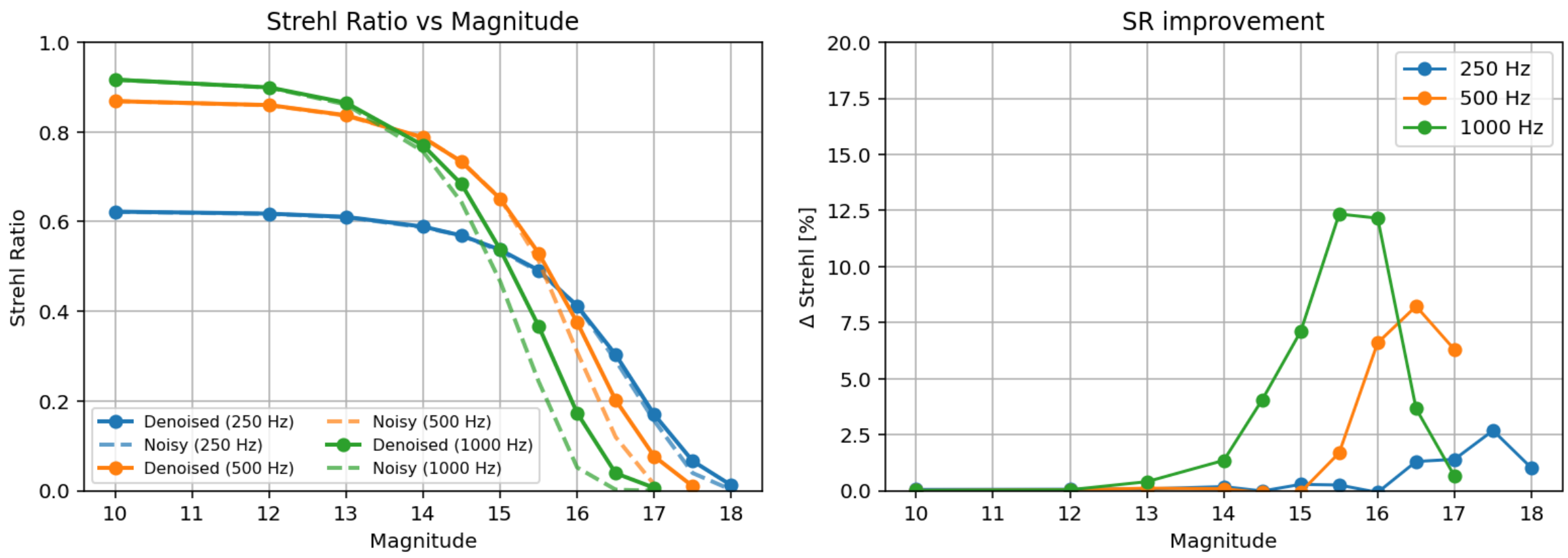

**Figure 4: Strehl Ratio as a function of magnitude (left). Improvement in SR ($SR_{denoised} - SR_{noisy}$) provided by the denoising algorithm (right).**

## 4.2 Hyperparameter tuning

The denoising threshold is controlled by the noise standard deviation estimate $\sigma$, which is obtained from the incoming PyWFS frame using the adaptive multilayer wavelet decomposition and PCA-based estimation procedure described in Section 2.1. In practice, this estimate must be scaled by a dimensionless factor $\alpha$, such that the effective threshold applied during collaborative filtering is proportional to $\sigma$. The optimal value of $\alpha$ varies with guide star magnitude, as shown in Figure 5.

The smooth dependence of performance on $\alpha$ follows directly from the properties of the underlying noise model. As the guide star flux decreases, the transition from a photon-noise-dominated regime to a read-noise-dominated regime occurs gradually, producing a continuous evolution in both the bias and variance of the PCA-based noise estimate. Because the noise statistics and the estimator response co-vary smoothly with signal level, the optimal scaling factor traces a regular, monotonic trend with magnitude rather than exhibiting discontinuous behaviour at any particular flux threshold. A practical consequence of this smoothness is that small deviations from the optimal α incur only minor perturbations to the effective threshold and therefore result in only marginal degradation in closed-loop Strehl ratio. This property is important for operational robustness: it implies that a look-up table of $\alpha$ values as a function of guide star magnitude -derived from the simulation results of Figure 5 - can be used to set the threshold in real time without requiring per-frame optimisation.

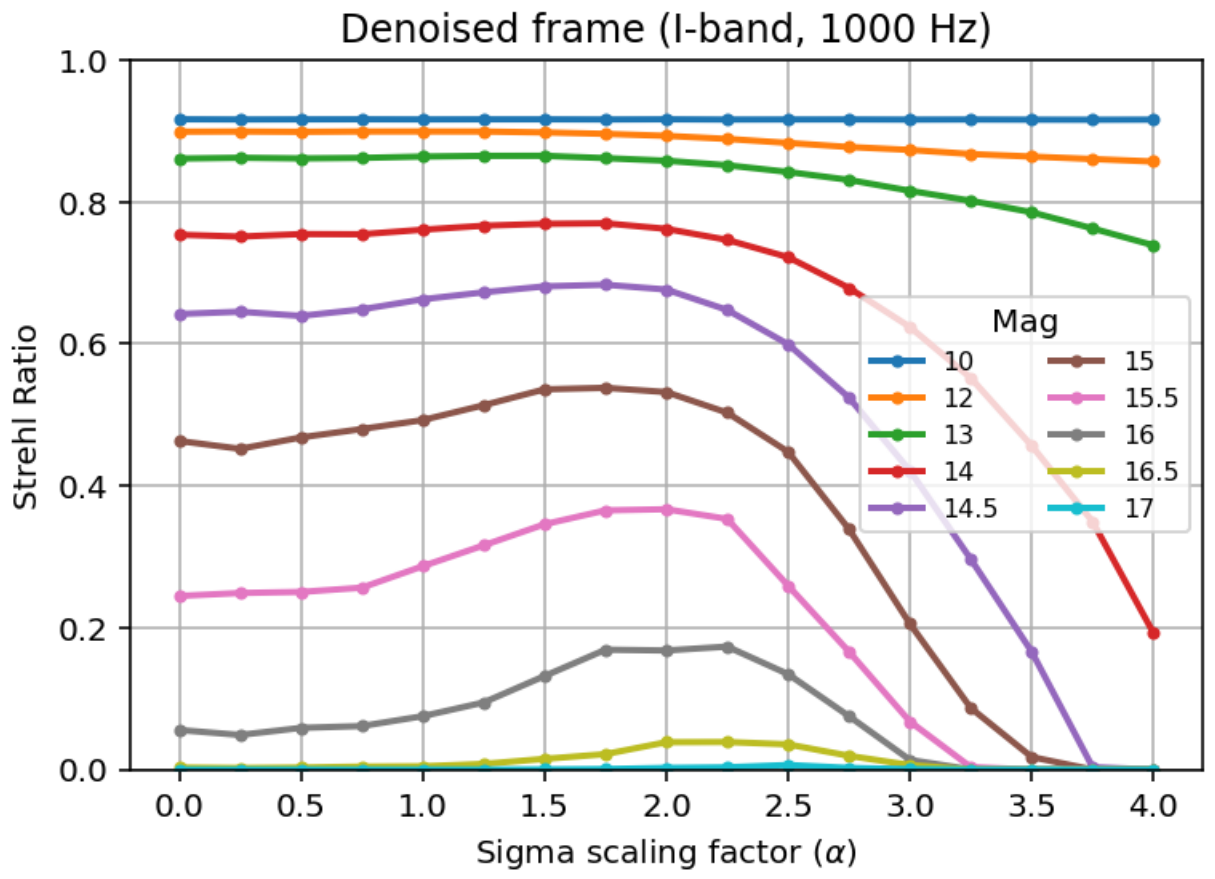

**Figure 5: Sigma scaling factor (α) as function stellar magnitude.**

### 4.3 Modal improvements

A comparison of the KL modal variances (Figure 6, left) shows a clear reduction for the higher-order modes. Significant improvements are also observed at low spatial orders, in particular for tip and tilt (Figure 6, right).
A limited number of modes, however, exhibit an increase in variance. These correspond to mid-spatial frequencies, with KL orders between 4 and 40. This suggests that the optimal α parameter is in fact modally dependent, and further optimisation of the denoising routine to account for this could improve the variance across all control modes (see below).

As noted previously, the number of controlled modes is unchanged between the noisy and denoised cases. Nevertheless, Figure 6 (left) suggests that the number of controlled modes could be increased by approximately 25% in the denoised case without compromising loop stability, potentially yielding further performance gains. The trend observed in Figure 6 (right) is consistent with the noise propagation behaviour (read-out and photon noise) reported in Chambouleyron et al. [8].

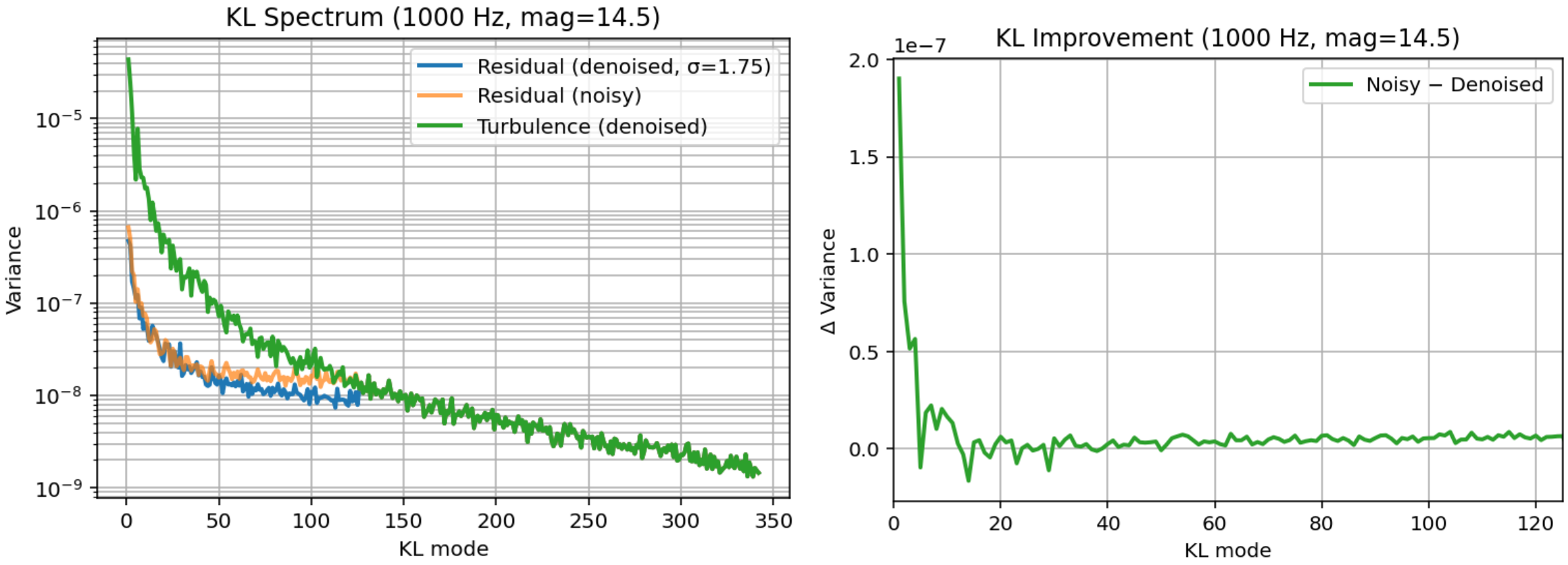

**Figure 6:Turbulent, noisy and denoised variance per KL mode (left). Difference of the KL variance between noisy and denoised PyWFS frames for the optimal alpha parameter of 1.75 (right).**

Figure 7 presents the ratio of the modal variances obtained from the noisy Karhunen–Loève (KL) coefficients to those obtained after denoising, i.e. $\sigma^2_{\text{noisy},k}/\sigma^2_{\text{denoised},k}$ for each mode $k$ and for different values of α. For the considered temporal frequency and stellar magnitude, the overall optimal regularisation parameter is found to be α = 1.75, for which the majority of modes exhibit ratios greater than unity, indicating a net variance reduction through the denoising process. A closer inspection reveals a non-uniform gain across modes as $\alpha$ is varied. This modal dependence suggests that a scalar global $\alpha$ might be suboptimal. Instead, a mode-dependent regularisation scheme $\alpha_k$ could optimise performance improvements across all modes. Furthermore, denoising may introduce unnecessary bias for certain modes. A hybrid reconstruction strategy in which denoising is selectively disabled for certain modes, while being retained for others might be optimal.

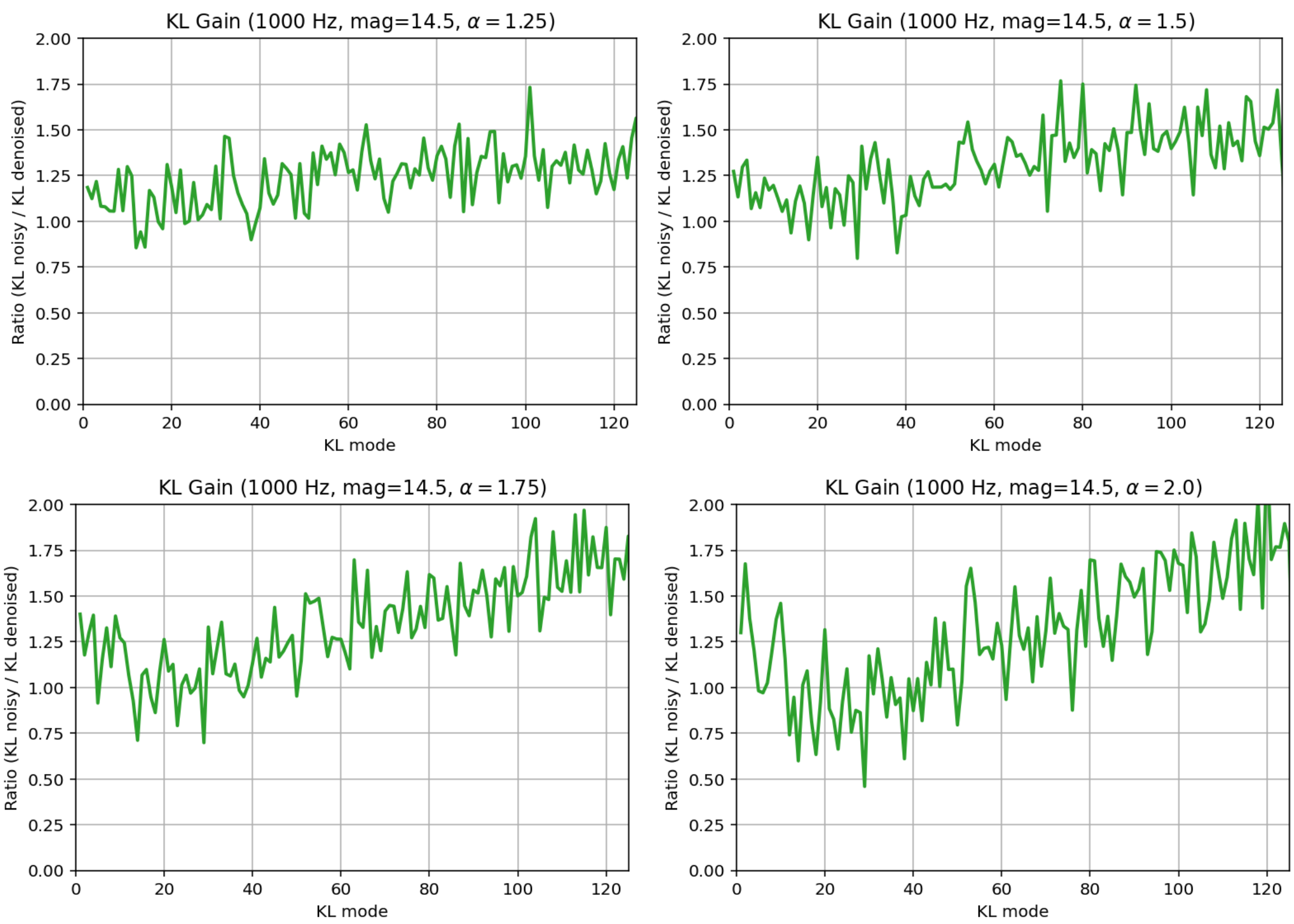

**Figure 7: Modal gain provided by the denoising algorithm as a function of the sigma scaling factor.**

We have already shown in Figure 5 that the optimal sigma scaling factor varies smoothly as a function of stellar magnitude. In addition, it also changes smoothly for a given KL mode. Figure 8 presents the maximum achievable gain when the scaling factor is optimally tuned independently for each KL mode. The results show that nearly all modes achieve a gain greater than unity. Only two modes (KL 18 and KL 38) fall slightly below this threshold, with gains close to 1, of approximately 0.95.

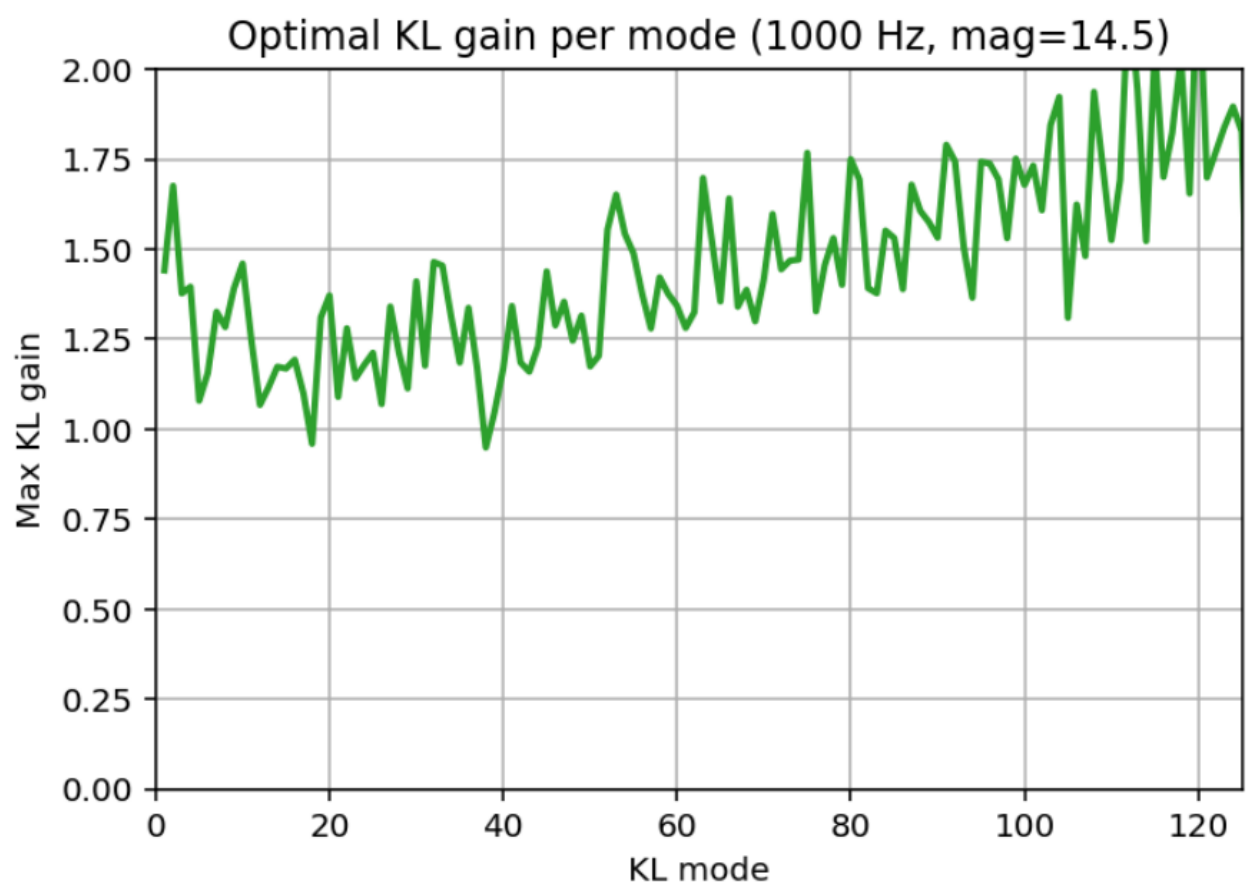

**Figure 8: Modal gain improvement provided by the denoising algorithm. The sigma scaling factor is selected independently for each KL mode.**

### 4.4 Impact on the PSF morphology

Application of the denoising algorithm increases the Strehl ratio from 64% to 68% (1 kHz sampling; magnitude 14.5). As expected, the overall PSF morphology is largely preserved (see Figure 9). The denoised PSF exhibits a slightly more compact coherent core, consistent with a modest reduction in residual wavefront error.

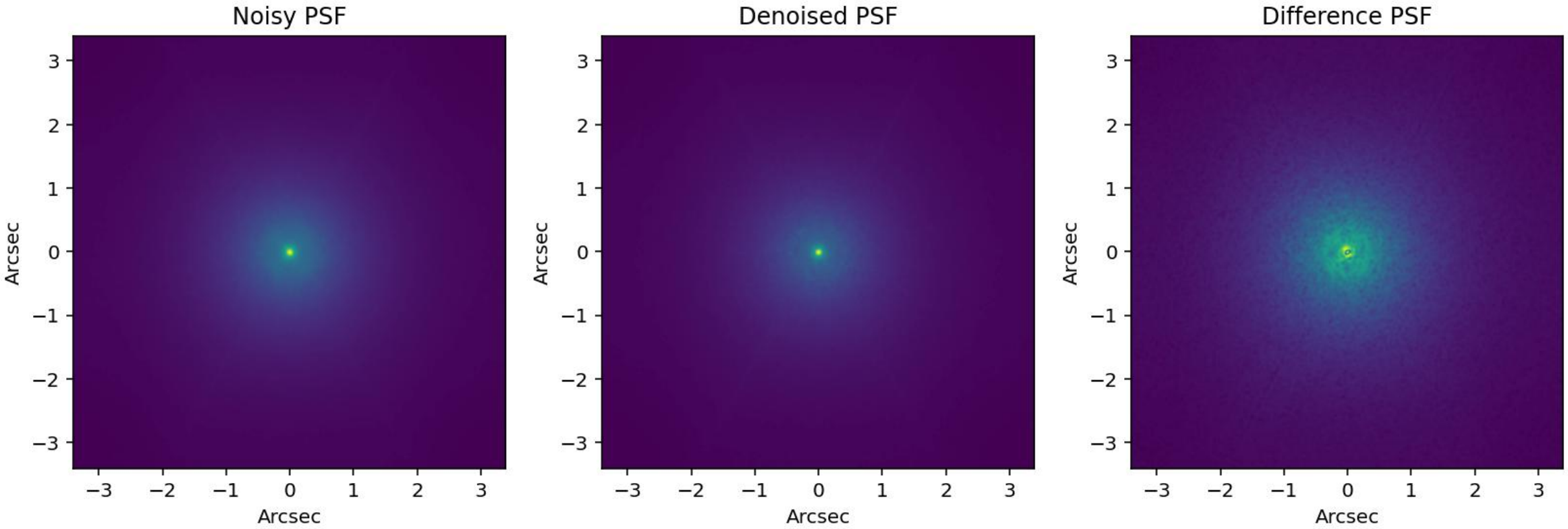

**Figure 9: Comparison of the noisy and denoised PyWFS PSFs in K-band, together with the difference map (right). All images are displayed using a cubic-root (power 1/3) intensity scaling to emphasise faint structures. 1.5 s integration time.**

Figure 10 presents radial cuts of the PSFs, illustrating the modest increase in Strehl ratio. The FWHM decreases from 0.077″ (1.35 λ/D) to 0.068″ (1.19 λ/D), corresponding to an 11.5% reduction. We expect that this large reduction is mainly due to the better correction of tip-tilt. Beyond these improvements, the denoising also leads to enhanced contrast, reflecting a more concentrated PSF core and reduced halo. This demonstrates the potential of the approach for high-contrast imaging applications.

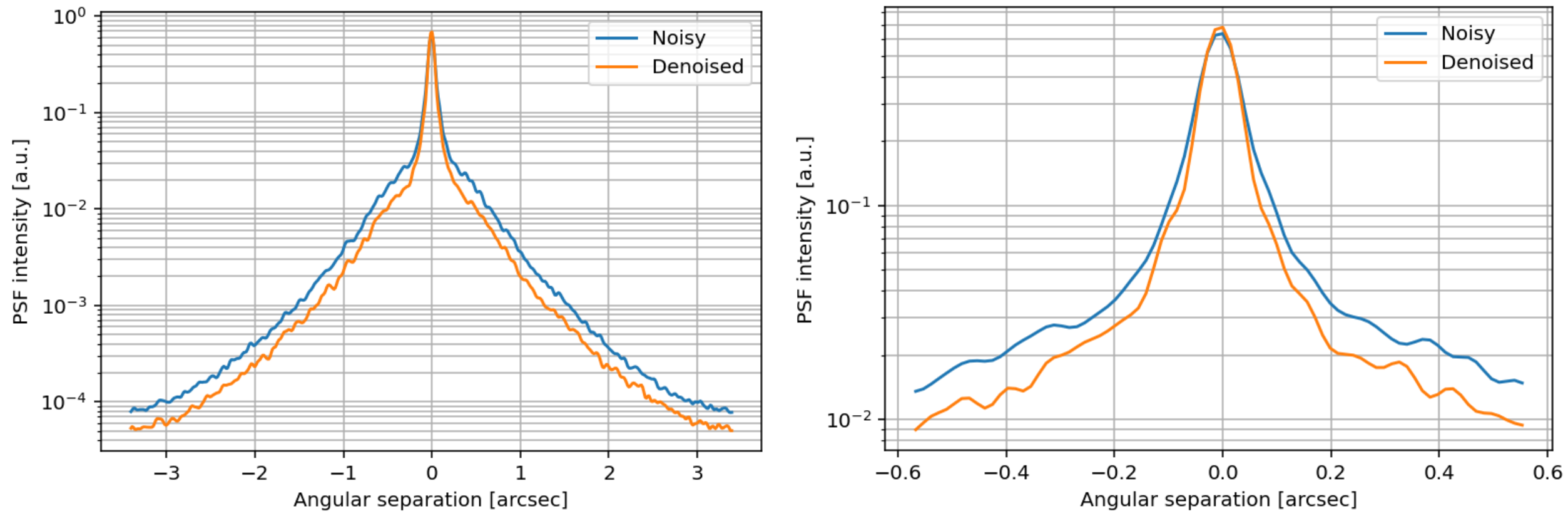

**Figure 10: Radial cut of the noisy PSF (SR = 63.7%) and denoised PSF (SR = 68.2%) in the K-band. (Left) zoom in the central area between approx. [-0.6, 0.6] arcsec.**

### 4.5 Impact of RON of Strehl Ratio

Figure 11 shows the variation of Strehl ratio as a function of the detector read-out noise (RON). For RON values of 1 and 3 $e^-$, the denoised cases exhibit a clear improvement in performance. Only marginal gains are observed for a very low read-out noise of RON = 0.1 $e^-$. Nevertheless, several aspects remain open for further optimisation see section 4.7. It is also important to note that these simulations assume a conventional CCD detector. They do not include the excess noise factor typically associated with electron-multiplying CCDs (EMCCDs) and Avalanche Photodiode Detectors (APD). In practice, while the effective read-out noise is reduced, the gain factor incurs excess noise which can reduce the effective signal-to-noise ratio. This effect should be included in future work.

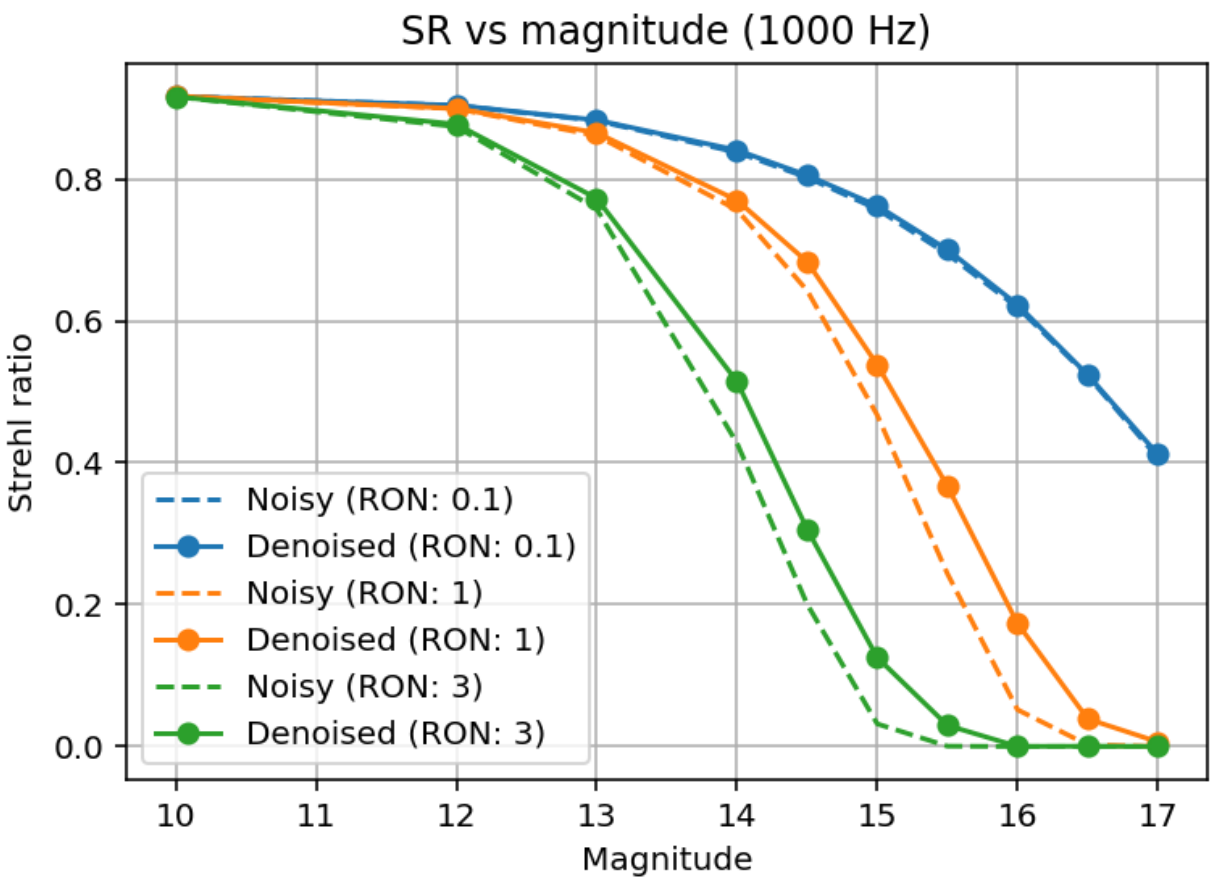

**Figure 11: Impact of read-out-noise on performance SR improvements.**

Figure 12 presents gain maps as a function of guide star magnitude and σ scaling factor for three detector read-out noise (RON) regimes (0.1, 1, and 3 e⁻) at a sampling frequency of 1 kHz. The colour scale represents the Strehl ratio gain relative to the noisy case. For RON = 3 e⁻, the analysis is restricted to magnitudes ≤15, beyond which performance degrades rapidly due to the lack of signal. In all cases, a clear ridge structure is observed, corresponding to the optimal σ as a function of magnitude. This ridge follows a smooth trend, reflecting the continuous evolution of the noise regime with decreasing flux.

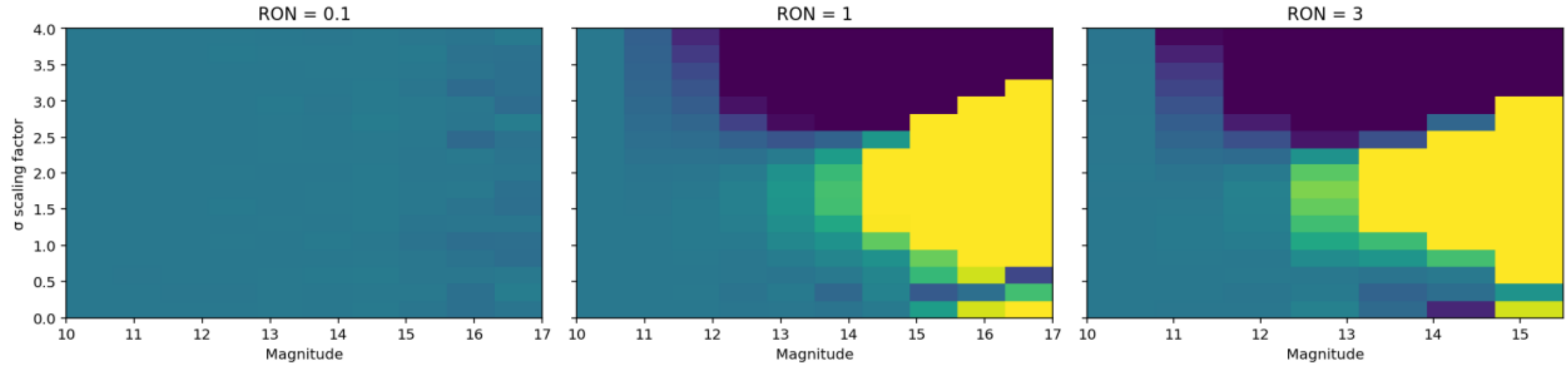

**Figure 12: Strehl gain ($SR_{denoised}$ / $SR_{noisy}$) vs magnitude and σ scaling parameter for RON = 0.1, 1, 3 e⁻ at 1 kHz.**

## 4.6 Impact of turbulence strength

In this section, we investigate the impact of turbulence strength on the performance gains achievable with denoising. Figure 13 presents results for three representative turbulence profiles: Median, JQ1, and JQ4, corresponding to Fried parameters $r_0$ of 0.15 m, 0.234 m, and 0.097 m (equivalent to seeing conditions of 0.67", 0.43", and 1.04"), respectively. The wind speed is identical to the median case, with a sampling frequency of 1 kHz, modulation set to 3λ/D and a detector read-out noise of 1 e⁻. The number of controlled KL modes is the same for all seeing conditions (see Table 2).

Under good seeing conditions (JQ1), the denoising algorithm yields significant performance improvements, with Strehl ratio gains reaching up to 25%. In contrast, for stronger turbulence (JQ4), only marginal improvements are observed, limited to approximately 5%. This suggests that further optimisation is required to fully realise the potential of the method in more challenging conditions. As expected, with improving seeing conditions, the AO system transitions from a regime dominated by fitting and temporal errors to one increasingly limited by wavefront sensor noise and its propagation through the reconstructor. Overall, these results highlight the effectiveness of the denoising approach in improving AO performance, particularly under favourable seeing conditions, and demonstrate its potential for high-contrast imaging applications.

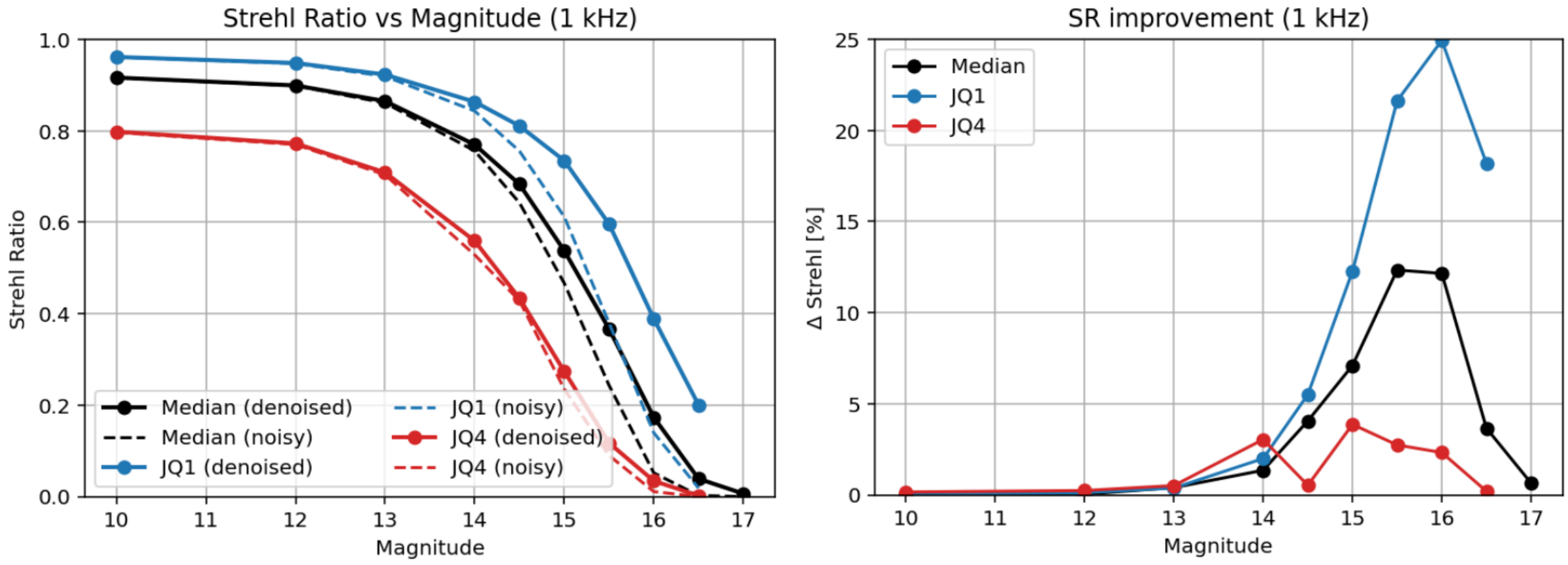

**Figure 13: [Left] SR for noisy and denoised cases function of magnitude and turbulence strength. [Right] SR improvement (i.e. difference between denoised and noisy). Modulation: 3λ/D.**

The optimal PyWFS modulation radius depends primarily on seeing conditions and flux (noise level). Smaller modulations are preferred in good seeing and high-flux regimes to maximise sensitivity, while larger modulations provide greater robustness under poorer seeing or noisier conditions. As modulation increases, PyWFS sensitivity decreases [8], increasing the impact of noise on correction performance.

Figure 14 shows the effect of modulation on Strehl ratio under JQ1 atmospheric conditions. As expected, smaller modulation improves performance, with a modulation of 1 being optimal in the non-denoised case. Applying denoising further improves performance, particularly at larger modulations where noise effects are stronger. Notably, the optimal modulation shifts from 1 to 2 with denoising, enabling operation in a more robust regime without sacrificing performance. Additionally, 30 KL modes are controlled here, compared to 20 in the previous case at the same magnitude (16). This demonstrates another benefit of denoising: it enables stable control of more modes, further improving correction quality.

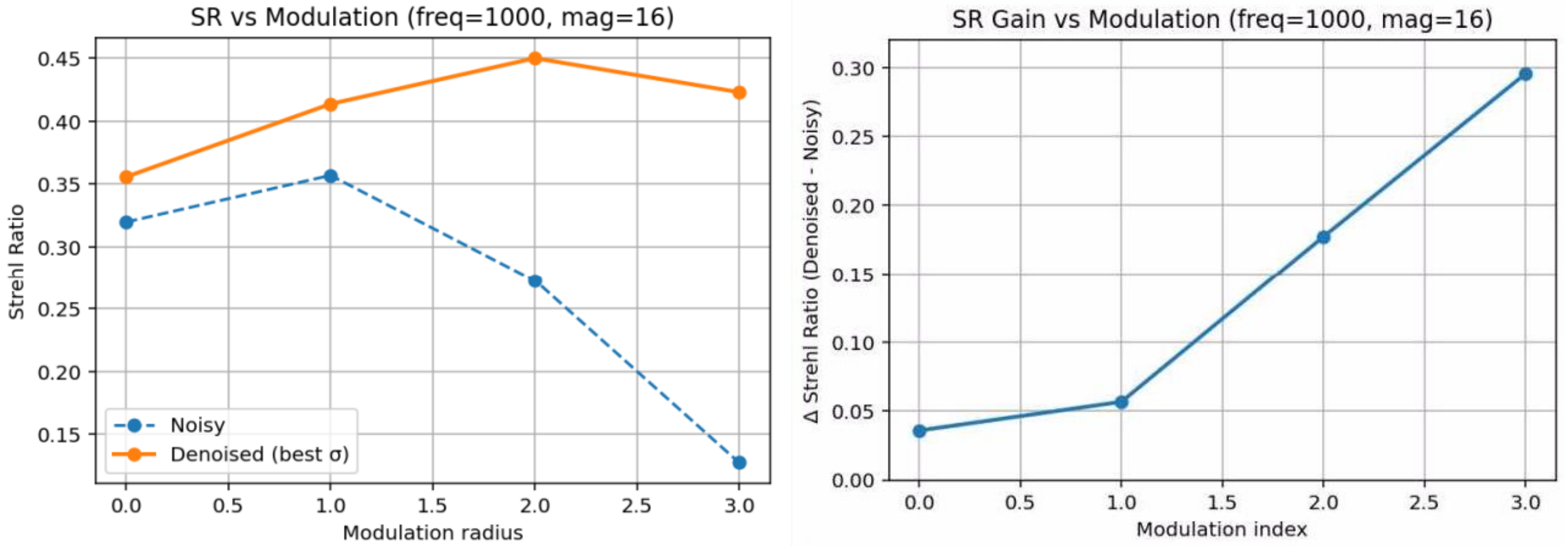

**Figure 14: Impact of modulation on Strehl ratio. Left: noisy and denoised PyWFS frame with optimal alpha. Right: SR gain for different modulation values. Number of controlled KL modes: 30.**

### 4.7 Complementary techniques and potential improvements

Several aspects of the current implementation offer clear opportunities for further optimisation:

1. **Number of controlled KL modes.** Figure 6 suggests that, at a magnitude of 14.5, the number of controlled modes could likely be increased from 125 (in the noisy case) to around 150 in the denoised case.

2. **Mode-dependent regularisation.** The α parameter could be optimised individually for each mode, rather than using a global value, to improve noise filtering and reconstruction fidelity (see Figure 8).
3. **Additional noise sources**. The current analysis does not include other contributors such as dark current or background noise, hot or dead pixels, which should be incorporated for a more realistic performance assessment.
4. **Optimal denoising algorithm for PyWFS**. The denoising algorithm as implemented applies a single global noise standard deviation $\sigma$ to the entire PyWFS frame and relies on a generic biorthogonal wavelet basis that was designed for natural photographic images. Both choices can be further optimized for the PyWFS context. The noise distribution on a PyWFS detector is spatially non-uniform - photon noise varies with the local intensity across the pupil, while read noise is uniform - so a per-pixel or per-sub-aperture noise map would provide a more accurate threshold at each spatial location. More fundamentally, the sparse representation underlying the collaborative filtering step determines how effectively signal and noise are separated in the transform domain. The four-pupil PyWFS intensity frame has a specific geometric structure - circular pupils with smooth, AO-corrected gradients and a characteristic quadrant symmetry - that a generic wavelet basis cannot capture. A representation tailored to this geometry, whether through a principal component basis derived from noiseless simulated frames, a dictionary learned from PyWFS-specific training data, or a structured basis that explicitly encodes the pupil geometry, would increase sparsity in the transform domain and sharpen the separation between signal and noise coefficients under thresholding. Taken together, these two refinements - spatially adaptive noise estimation and a PyWFS-optimized sparse basis - define the most technically substantive path toward an optimal denoising algorithm specifically designed for wavefront sensing.

Some systems utilise hardware binning of the detector pixels in faint guide star regimes, reducing the number of modes controlled but increasing the total SNR where readout noise dominates the measurement. Additional improvements may be also achieved through optimisation of the loop gain for each mode and through the use of regularised or MMSE reconstructors that explicitly account for measurement noise. We note, however, that these techniques are largely independent of the denoising methodology and would be expected to improve both noisy and denoised cases. As such, they do not affect the generality of the conclusions presented here. Instead, they should be viewed as complementary approaches that could be combined with denoising to further enhance AO performance in photon-limited regimes.

## 5 CONCLUSION AND NEXT STEPS

Image-domain denoising provides a simple and effective means to improve the sensitivity and robustness of adaptive optics (AO) systems in noise-limited regimes relevant to next-generation instrumentation. Applied to pyramid wavefront sensor (PyWFS) images, full-frame denoising yields measurable improvements in Strehl ratio and PSF quality under low-flux conditions. Furthermore, the denoised case supports control of a larger number of KL modes, enabling additional performance gains.

The impact of the method depends on detector characteristics, with diminishing returns in the low read-noise regime - an area we will investigate in more detail in our future work. This makes the approach particularly attractive for high-noise applications, such as infrared wavefront sensing, as well as for cost-effective AO systems and experimental platforms. It also remains relevant for high-performance systems (e.g. PCS), where gains are amplified under favourable seeing conditions.

Future work will focus on improving noise modelling and extending the applicability of the method. In particular, the current use of a global noise estimate (σ) scaled by a single factor (α) does not capture the spatial variability of PyWFS noise, which arises from signal-dependent photon statistics and uniform read noise. Introducing spatially varying noise maps, defined per pixel or sub-aperture, is expected to improve denoising performance, especially in low-flux regions and at pupil edges. In parallel, modal analysis indicates that optimal regularisation varies across modes; a mode-dependent scaling applied in slope space can therefore outperform a global parameter. These spatial and modal refinements represent the most immediate path to improved performance.

Further gains may be achieved by developing sparse representations tailored to PyWFS images. The current use of generic wavelet bases does not fully exploit the structured nature of the data, characterised by four pupils, smooth intensity distributions, and quadrant symmetry. Bases derived from simulated data (e.g. PCA), learned dictionaries, or analytically constructed transforms could provide more compact representations, improving the separation of signal and noise during thresholding.

Finally, real-time implementation at kHz frame rates is feasible with appropriate optimisation. The dominant computational cost - the 3D wavelet transforms over grouped patches - is well suited to GPU acceleration. In addition, the noise estimation stage can be updated at a reduced cadence, as the underlying parameters evolve slowly compared to the AO loop rate. This decoupling significantly reduces computational load and supports practical real-time deployment on operational AO systems.

## ACKNOWLEDGEMENTS

This research is funded by the STFC's Centre for Innovation (CfI).